\documentclass{ws-jnopm}
\usepackage{graphicx}
\usepackage{epstopdf} 
\usepackage{color}    

\newcommand{\smu}{Southern Methodist University}
\newcommand{\pt}{$\mathcal{PT}$}

\renewcommand\Im{\operatorname{Im}}

\begin{document}
 
\markboth{C. Castro-Castro et al.}{Light dynamics in nonlinear trimers and twisted multicore fibers}

\catchline{}{}{}{}{}

\title{LIGHT DYNAMICS IN NONLINEAR TRIMERS AND TWISTED\\
MULTICORE FIBERS
}

\author{CLAUDIA CASTRO-CASTRO\footnote{Corresponding author.}} 

\address{Department of Mathematics, \smu,
Dallas, Texas 75275, USA\\
ccastrocastr@smu.edu}

\author{YANNAN SHEN}

\address{Department of Mathematics, University of California Northridge, California 91330, USA\\
yannan.shen@csun.edu}

\author{GOWRI SRINIVASAN}
\address{Theoretical Division, Los Alamos National Laboratory, Los Alamos, NM 87544}%

\author{ALEJANDRO B. ACEVES} 
\address{Department of Mathematics, \smu, Dallas, Texas 75275, USA}

\author{PANAYOTIS G. KEVREKIDIS}
\address{Department of Mathematics and Statistics, University of Massachusetts, Amherst, MA 01003-4515, USA}%

\maketitle

\begin{history}
\received{(Day Month Year)}
\end{history}

\begin{abstract}
Novel photonic structures such as multi-core fibers and graphene based arrays present unique opportunities 
to manipulate and control the propagation of light. Here we discuss nonlinear dynamics for 
structures with a few (2 to 6) elements for which linear and nonlinear properties can be tuned. 
Specifically we show how nonlinearity, coupling, and parity-time (\pt) 
symmetric gain/loss relate to existence, stability and in general, dynamical properties of nonlinear 
optical modes. The main emphasis of our presentation will be on systems with few degrees of freedom, 
most notably couplers, trimers and generalizations thereof to systems with 6 nodes.

\end{abstract}


\section*{Introduction}

Advancements in photonics are driven by the possibilities they present to all optical devices. 
Complex array geometries and the use of novel materials such as graphene and metamaterials 
for which one can tune properties such as coupling, nonlinearity and gain/loss translate into 
controlled dynamics and bifurcations of interest to logical devices (switches, transistors, logic gates),
next generation lasers and new formats for optical communications.\cite{0-suchkov2016nonlinear}

Here, we follow the vein of the work of Bender {\it et al}.\cite{1-bender2007making} who considered whether a quantum mechanical 
Hamiltonian system with a complex potential can have a real spectrum. It is the case that a necessary condition for the
spectrum to be real is that the Hamiltonian of the Schr\"odinger has parity $\mathcal{P}$ and time $\mathcal{T}$  symmetry (\pt) which is
true when the potential satisfies the condition $V(x)=V^*(-x)$.\cite{2-miroshnichenko2011nonlinearly}
This has led to the identification of a broad range of special classes of \pt-symmetric Hamiltonians which 
possess a real eigenspectrum and \pt-symmetric eigenstates. Perhaps more relevant is that if the strength of the 
odd imaginary part of the potential is parametrized, one typically finds that at a certain parameter value the 
spectrum becomes complex. 
Whenever the eigenfrequencies remain real, we refer to this case as $unbroken$ \pt-symmetry. 
If they become complex, one typically refers to this as the regime where \pt-symmetry has been {\em broken}. This work originally 
presented in the quantum mechanical framework can be naturally extended to the optical regime. 
In fact, it was in the latter setting that \pt-symmetric systems were first experimentally realized 
in the form of dimers,\cite{3-ruter2010observation,4-guo2009observation}  (i.e., 
optical waveguide couplers)  and more recently in the realm of
full one-dimensional arrays.\cite{5-wimmer2015observation} 
Novel photonic structures can be conceived with \pt-symmetry properties, which translate to having a 
real index of refraction profile that is even in space, whereas the gain/loss imaginary part is odd. 
Manipulating light propagation under \pt-symmetric structures and what broken symmetry entails has been the 
focus of intense research activity in recent time.\cite{6-nixon2016nonlinear,7-sukhorukov2012nonlocality}

In this paper we present results on nonlinear \pt-symmetric discrete arrays focusing especially on the case of a few lattice nodes. 
In such photonic systems there are four principal parameters that characterize the type of nonlinear modes,
their propagation features and possible bifurcations: nearest neighbor coupling, individual propagation constants, 
nonlinear coefficient and \pt gain-loss coefficients. Altogether, dynamics in nonlinear photonic arrays is currently a rich field 
of active research. Here we present results for three configurations: a dimer, a trimer and a most 
recently considered setting of twisted multicore fibers.\cite{8-longhiPTphase2016}

The paper is organized as follows. In Section \ref{sec:PTcoupler}, we introduce the coupled \pt-symmetric 
nonlinear Schr\"odinger (NLS) array with two and three sites. We will refer to these cases as coupler and trimer respectively. 
While couplers and trimers have been thoroughly studied, this section primarily serves to illustrate
\pt dynamics in the simplest case.
Section \ref{sec:twisted-multicore} presents our recent work which extends that 
of Ref.~\refcite{8-longhiPTphase2016} by inclusion of nonlinear effects. As a way of comparison, we present numerical results for a conservative 
set up, and of cases with gain or loss in the individual cores. Section \ref{sec:discussion} summarizes our results
and presents some possibilities for further study.

\section{\pt-symmetric coupler and trimer}\label{sec:PTcoupler}

\begin{figure}[ht!]
 \centering
 \includegraphics[width=0.6\textwidth]{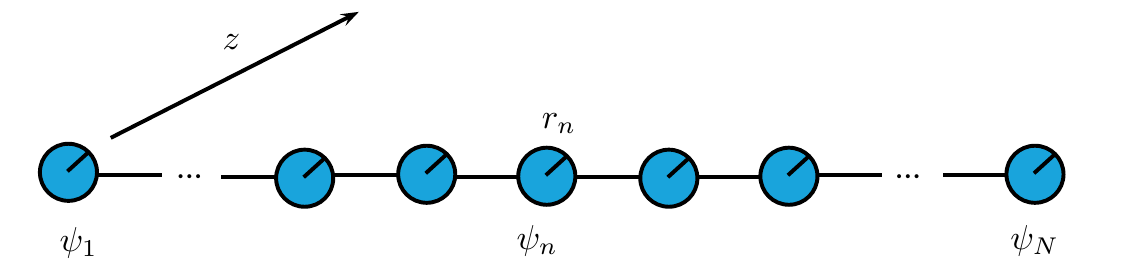}
 \caption{Diagramatic representation of a planar optical waveguide array. $r_n$ is the radius of each particular fiber.}
 \label{fig:1dDNLS_diag}
\end{figure}

Consider the propagation dynamics in a discrete array of $N$ optical fibers in Kerr media described by the one-dimensional
discrete Nonlinear Schr\"odinger Equation (DNLSE):
\begin{equation}\label{eq:DNLSE}
i \dot{\psi_n} =  \epsilon_n \psi_n + \sigma|\psi_n|^2\psi_n - c(\psi_{n+1}+\psi_{n-1}),
\end{equation}
where $n$ is the waveguide number, the dot denotes derivative with respect to the direction of
propagation $z$,  $\sigma$  is the strength of the Kerr-nonlinearity, $c$ is the uniform strength of the nearest neighbor 
coupling, $\epsilon_n$ represents the on-site refractive-index profile (Fig. \ref{fig:1dDNLS_diag}). 
The Hamiltonian that gives rise to the equations of motion is given by the equation 
\begin{equation}\label{eq:hamiltonian}
H_D = \sum_n \epsilon_n|\psi_n|^2 + \frac{\sigma}{2}|\psi_n|^4 - c\left(\psi_{n+1}\psi_n^*+\psi^*_{n+1}\psi_n  \right).
\end{equation}
We start considering an ideal configuration of identical optical wave guides, where $\epsilon_n$ is a fixed constant $\epsilon$.

\subsection{The coupler}

We explore the eigenspectrum associated with a structure which has complex valued 
Hamiltonian with \pt-symmetry composed of two waveguides, one with loss at location
$n=1$ and the other with gain at the adjacent waveguide $n=2$. That is
$\epsilon_n =  \alpha_n + i \gamma\beta_n,$
where $\gamma$ defines the rate of gain/loss at each waveguide, 
and $\alpha_n$ and $\beta_n$ satisfy the even and odd conditions
$\alpha_2 = \alpha_{1}$ and $\beta_2 = -\beta_{1}$. 
The propagation equations for this case can be written as 
\begin{subequations}
\label{eq:linearcase1}
\begin{align}
i\dot{\psi_{1}}	=&\delta \sigma|\psi_{1}|^2\psi_{1}-c\psi_{2}+\alpha\psi_{1}+i\gamma\beta\psi_{1},\\
i\dot{\psi_{2}}	=&\delta \sigma|\psi_{2}|^2\psi_{2}-c\psi_{1}+\alpha\psi_{2}-i\gamma\beta\psi_{2}.
\end{align}
\end{subequations}

It is well known based on the work of Ramezani {\it et al.}\cite{9-ramezani2010unidirectional} that the coupler is an
integrable dynamical system due to the existence of two conserved quantities in the form of  
$C^2=4\left[\delta^2 \sigma^2|\psi_1|^2|\psi_2|^2-\delta \sigma\left(\psi_1^*\psi_2+\psi_1\psi_2^* \right) +1\right],$ and 
$J=|\psi_1|^2+|\psi_2|^2+(2\beta/\delta \sigma)\sin^{-1}\left\lbrace\left[\delta \sigma\left(\psi_1^*\psi_2+\psi_1\psi_2^*\right)-2\right]/C\right\rbrace.$ 
Moreover, the integrability has been illustrated in a complementary fashion by Barashenkov\cite{10-barashenkov2014hamiltonian}
who showed that upon suitable coordinate transformations, the system can be turned into a (conservative) Hamiltonian one.

In the weakly nonlinear regime, we have developed a modulation theory for the coupler problem.
Observe that the linear model has eigenvalues $\lambda_{1,2}=\alpha\pm \sqrt{c^2-\beta^2\gamma^2}$ which are 
real as long as the gain/loss parameter $\gamma$ is smaller than a critical value
$\gamma_c=c/\beta$. The respective eigenvectors are 
\begin{center}
$\begin{pmatrix}
-e^{i\theta}\\
1 
\end{pmatrix}$ and $ \begin{pmatrix}
e^{-i\theta}\\
1
\end{pmatrix},$
\end{center}
where $\theta$ is defined such that $\sin{\theta}=\beta \gamma/c$.

In the case of weak nonlinearity, the evolution equations read
\begin{equation}\label{eq:NLPTcoupler}
i\dot{\psi}=VDV^{-1}\psi+\delta\sigma N(\psi),
\end{equation}
where $\psi=(\psi_1,\psi_2)^T$, $VDV^{-1}$ is the diagonal decomposition of the block matrix associated with
the linear eigenvalue problem, $\delta$ is a small parameter, and the nonlinear term is given by
$$N(\psi)=\begin{bmatrix}|\psi_{1}|^{2}\psi_{1}\\
|\psi_{2}|^{2}\psi_{2}
\end{bmatrix}=\begin{bmatrix}|\psi_{1}|^{2} & 0\\
0 & |\psi_{2}|^{2}
\end{bmatrix}\psi.$$
Multiplying by $V^{-1}$ from the left on both sides of (\ref{eq:NLPTcoupler}) and defining $\psi=V\phi$, 
we can rewrite Eq. (\ref{eq:NLPTcoupler})  as
\begin{equation} \label{eq:NLPTcouplerPhi}
i\dot{\phi}=D\phi+\delta\sigma V^{-1}N(V\phi).
\end{equation}
From the linear solution of the previous equation in terms of $\phi$, we can write the solution to the
initial value problem as $\phi(z)=e^{-iDz}C$, where $C=C(z)$
\begin{equation}
\label{eq:NLPTcouplerPhiFull}
i\frac{dC}{dz}=\delta\sigma e^{iDz}V^{-1}\begin{bmatrix}|\psi_{1}|^{2} & 0\\
0 & |\psi_{2}|^{2}
\end{bmatrix}Ve^{-iDz}C
\end{equation}

For brevity, we restrict the model to two scenarios: \emph{a)}  we assume \pt-symmetry as small perturbation of the 
original system, $\alpha=1$, $\beta=\delta$, where $\delta$ is a small parameter, $\delta\ll1$. 
In the second case \emph{b)}  we consider the regime near the 
\pt-symmetry transition region; that is when $\gamma$ is similar to $c$.
We aim to derive amplitude equations as slowly varying functions of time.
In the first case \emph{a)}, the right hand side of (\ref{eq:NLPTcouplerPhiFull}) has oscillating terms with frequencies 
proportional to $\lambda_1-\lambda_2$, which can also be writen as $2c\cos{\theta}$, where $\sin{\theta}=\delta\gamma/c$. 
These terms represent a fast osci\-llating evolution when $\delta$ is a small parameter. We thus perform an averaging and keep only the slow oscillating (near resonant) terms.
Introducing the scale variable $Z=\delta z$, we can represent the amplitudes as slowly varying functions of $Z$: 
$C_{k}(z;\delta)=\tilde{C}_{k}(Z)$ whose evolution is given by
\begin{subequations}
\label{eq:PTcoupler}
\begin{align}
i\frac{d\tilde{C}_{1}}{dZ}&=\sigma\left(|\tilde{C}_{1}|^{2}\tilde{C}_{1}+2|\tilde{C}_{2}|^{2}\tilde{C}_{1}\right),\\
i\frac{d\tilde{C}_{2}}{dZ}&=\sigma\left(|\tilde{C}_{2}|^{2}\tilde{C}_{2}+2|\tilde{C}_{1}|^{2}\tilde{C}_{2}\right).
\end{align}
\end{subequations}

We note that system (\ref{eq:PTcoupler}) can be solved analytically given the existence of two conserved quantities, 
$|\tilde{C_1}|^2$ and $|\tilde{C_2}|^2$. The resulting dynamics is a trivial phase modulation to the linear modes.

A more interesting case is \emph{b)}, representing the regime near the transition region. We can no longer average the 
terms in (\ref{eq:NLPTcouplerPhiFull}) with frequency proportional 
to $\lambda_{1}-\lambda_{2}=2c\cos{\tilde{\theta}}$, where $\sin{\tilde{\theta}}=\gamma/c$, which is close to zero.
Let $k=(\lambda_{1}-\lambda_{2})/2\delta$, with $k=\mathcal{{O}}(1)$. 
Introducing again the slow scale variable $Z=\delta z$ and using the definition of the parameter $k$, the system of amplitude equations 
can be written as
\begin{subequations}
\label{eq:PTcouplerMod2}
\begin{align}
i\frac{d\tilde{C_{1}}}{dZ} & = \sigma\left(|\tilde{C_{1}|}{}^{2}\tilde{C_{1}}+2|\tilde{C_{2}}|{}^{2}\tilde{C_{1}}+2e^{-2ikZ}\tilde{C_{1}^{2}}\tilde{C_{2}^{*}}\right)-\gamma e^{4ikZ}\tilde{C_{1}^{*}}\tilde{C_{2}^{2}}, \label{eq:PTcouplerMod2a}\\
i\frac{d\tilde{C_{2}}}{dZ} & = \sigma\left(|\tilde{C_{2}|}{}^{2}\tilde{C_{2}}+2|\tilde{C_{1}}|{}^{2}\tilde{C_{2}}+2e^{2ikZ}\tilde{C_{1}^{*}}\tilde{C_{2}^{2}}\right)-\gamma e^{-4ikZ}\tilde{C_{1}^{2}}\tilde{C_{2}^{*}}.\label{eq:PTcouplerMod2b}
\end{align}
\end{subequations}

One finds that the squared amplitudes satisfy:
\begin{subequations}
\label{eq:PTcouplerIntensities}
\begin{align} 
\frac{d|\tilde{C}_{1}|^{2}}{dZ} = & 2\sigma\left[2|\tilde{C}_{1}|^{2}\Im{\left(e^{-2ikZ}\tilde{C}_{1}\tilde{C}_{2}^{*}\right)}+\Im{\left(e^{-4ikZ}\tilde{C}_{1}^{2}\tilde{C}_{2}^{*2}\right)}\right],\label{eq:PTcouplerIntensities_a}\\
\frac{d|\tilde{C}_{2}|^{2}}{dZ} = & 2\sigma\left[2|\tilde{C}_{2}|^{2}\Im{\left(e^{2ikZ}\tilde{C}_{1}^{*}\tilde{C}_{2}\right)}+\Im{\left(e^{4ikZ}\tilde{C}_{1}^{*2}\tilde{C}_{2}^{2}\right)}\right],\label{eq:PTcouplerIntensities_b}
\end{align}
\end{subequations}
\noindent
which in particular shows that the system is no longer integrable, but the amplitudes do not live in a circle. Future work will expand on the full analysis of the 
weakly nonlinear regime, in particular the case near transition. 

\subsection{The trimer}

For the trimer, here we only address the question of whether a region of unbroken \pt-symmetry persists in the linear case for 
on-site refractive-index profile,
\begin{center} $\epsilon_n =  \alpha_n + i \delta \gamma\beta_n,$\end{center}
where $\delta$ is a small parameter, $\gamma$ defines the rate of gain/loss at waveguides, 
and $\alpha_n$ and $\beta_n$ satisfy the even and odd conditions
$\alpha_n = \alpha_{-n}$ and $\beta_n = -\beta_{-n}$. 
For an open linear trimer with an active, a lossy, and a conservative
waveguide, we seek solutions of the form $\psi_n = A_n e^{-i\lambda z}$, 
then $(A_1,A_2,A_3)$ satisfies
\begin{eqnarray*}\label{eq:DNLSElinear}
\lambda A_1 &=&  \epsilon_1 A_1 -cA_{2}\\
\lambda A_2 &=&  \epsilon_2 A_2 -c(A_{3}+A_{1})\\
\lambda A_3 &=&  \epsilon_3 A_3 -cA_{2}
\end{eqnarray*}
For  the \pt-symmetric case of a homogeneous array with $\alpha_n = \beta_n = 1$ the characteristic polynomial 
of the associated matrix problem is
$$p(c,\gamma)=-2 c^2 \lambda +2 c^2 +\lambda ^3-2 \lambda ^2+\lambda+\lambda \gamma^2.$$
In order to explore whether a region of transition (real to complex eigenvalues) exists, we check the sign of 
the discriminant of the third degree polynomial associated with the nontrivial roots of $p(c,\gamma)$ is
\begin{multline}\label{discriminant}
\Delta = -108 c^4-72 c^2 \left(-2 c^2+\gamma^2+1\right)-4 \left(-2
   c^2+\gamma^2+1\right)^3\\+4 \left(-2 c^2+\gamma^2+1\right)^2+64 c^2 
\end{multline}

\begin{figure}[ht!]
 \centering
 \includegraphics[width=0.40\textwidth]{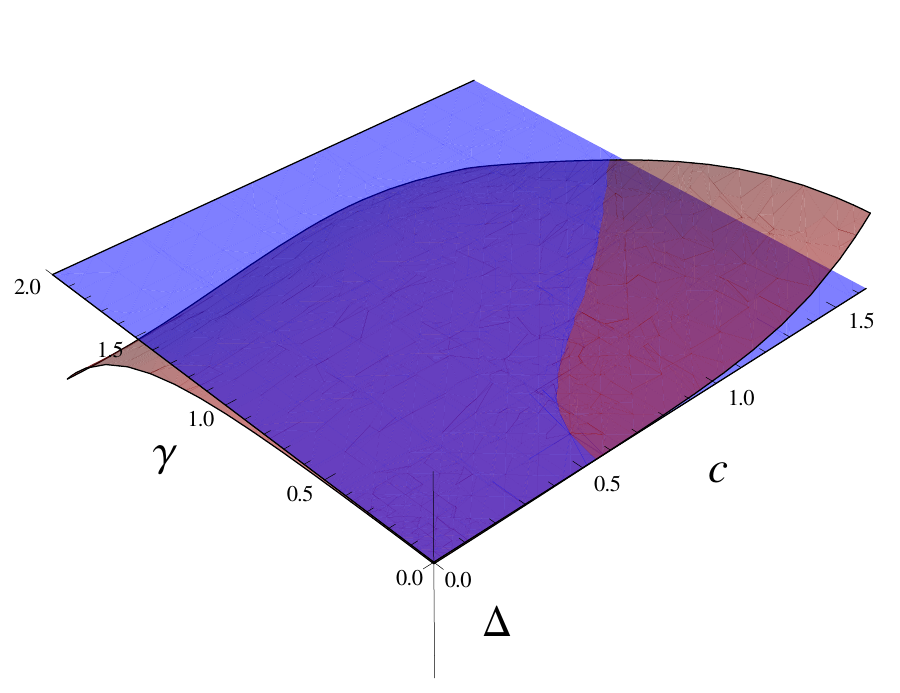}
  \includegraphics[width=0.30\textwidth]{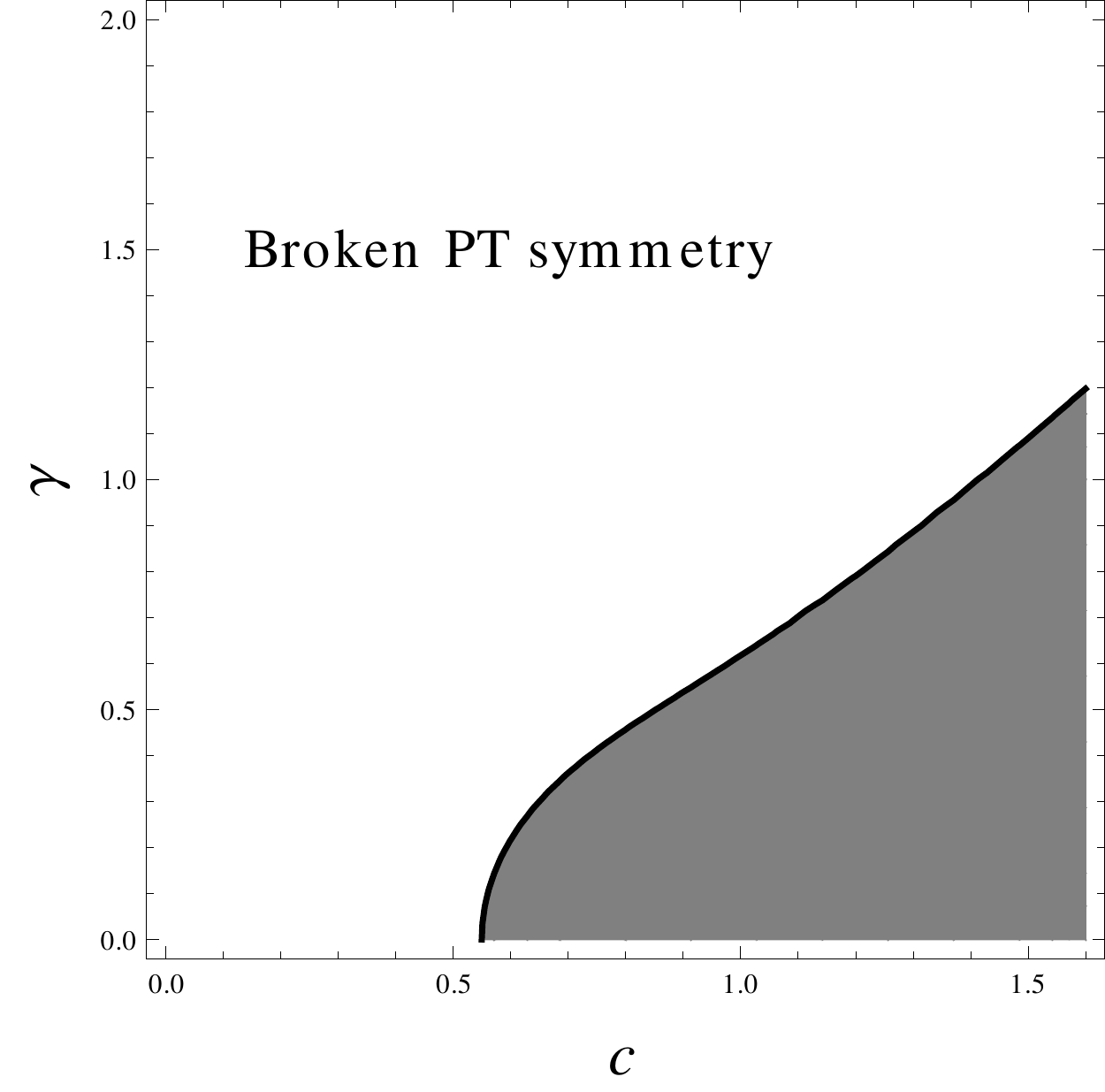}
 \caption{(Color online) Discriminant as function of coupling $c$ and gain/loss rate $\gamma$. The shadded region in the right 
 panel shows the region of unbroken \pt-symmetry in the $(c,\gamma)$ plane.}
\label{fig:discriminantPT}
\end{figure}

\noindent
In Figure \ref{fig:discriminantPT} we plot the discriminant as a function of $c$ and $\gamma$ as a red surface and 
the plane $\Delta =0$ as a blue surface. We observe that there exists a region of the $(c,\gamma)$ plane where 
real eigenfrequencies exist whenever the discriminant is positive. We also note a transition region from broken to unbroken 
\pt-symmetry, or viceversa, is represented as the intersection curve of the red and blue surfaces.

Future work will explore the dynamics for parameters in the vicinity of $\Delta(\alpha, \gamma,c)=0$ which will 
include weak nonlinear effects.

\section{Twisted multicore fibers} \label{sec:twisted-multicore}

\begin{figure}[ht!]
 \centering
 \includegraphics[width=0.3\textwidth]{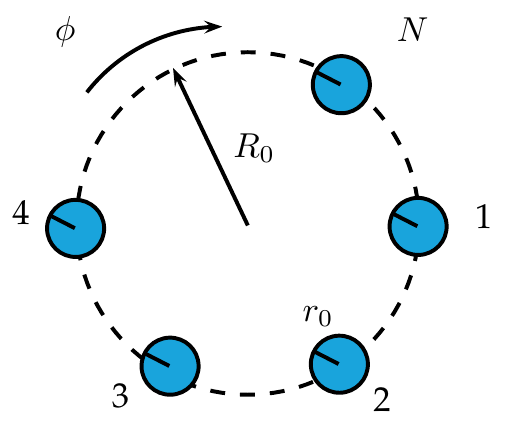}
 \caption{Diagramatic representation of a multi-core fiber. $R_0$ radius of the circular fiber and $r_0$ is the radius of each particular fiber.
 The fiber is twisted  along the propagation direction $z$.}
 \label{fig:1dDNLS_diag_multicore}
\end{figure}

In this section, we consider the beam propagation dynamics in a discrete multi-core fiber of $N$ optical waveguides arranged equally spaced
in a ring of radius $R_{0}$ where each core has radius $r_0$ (Figure \ref{fig:1dDNLS_diag_multicore}), described by the coupled-mode equations (CMEs)
\begin{equation}
i\frac{{dc_{n}}}{dz}=k\left(e^{-i\phi}c_{n+1}+e^{i\phi}c_{n-1}\right)+i\gamma_{n}c_{n}+\delta d|c_{n}|^{2}c_{n}. 
\label{eq:twisted-PT-DNLS}
\end{equation}
Here $n=1,2,...,N$ is the waveguide (site) number, $c_{n}$ represents complex-valued amplitudes that depend on the 
direction of propagation variable $z\in\mathbb{R}$, $k$ is the uniform strength of the nearest neighbor coupling, $\gamma_{n}$
is the optical gain or loss rate at site $n$. The fiber is considered to be twisted along the propagation direction $z$
with a twist rate $\varepsilon=2\pi/\Lambda$, $\Lambda$ is the spatial period. 
Recent work by Longhi\cite{8-longhiPTphase2016} only considers the linear $\delta d=0$ case where he shows that
transition from unbroken to broken PT phases can be conveniently
controlled by a suitable geometric twist of the fiber.
The fiber twist which is responsible for the introduction of additional phase terms in the propagating fibre modes 
$e^{\pm i\phi}$, $\phi$ is defined as the Peierls phase,\cite{11-longhiLightTransfer2007} \ $\phi=4\pi^{2}\varepsilon n_{s}r_{0}^{2}/N\lambda$, 
where $n_{s}$ is the substrate (cladding) refractive index, $r_{0}\simeq R_{0}$ is an effective ring radius, $\lambda$
is the wavelength of the propagating field. As such, 
in Ref. \refcite{11-longhiLightTransfer2007} the author finds that for three fibers with gain loss $\gamma_1=0, \gamma_2=-\gamma_3=\gamma$, 
the unbroken parity time phase is obtained when the condition \ 
$\left(\gamma/k\right)^2 < 3\sqrt{1-\cos^{2/3}(3\phi)}$ is satisfied. We want to extend this work to larger arrays and account for nonlinear effects.
  
\noindent
From now on, we will refer to equation 
(\ref{eq:twisted-PT-DNLS}) as twisted \pt-DNLS equation, which can be rewriten as
\begin{equation}
i\dot{\mathbf{c}}=\left(k\mathcal{\mathbb{\mathcal{K}}}+iG\right)\mathbf{c}+\delta d N\left(\mathbf{c}\right)\mathbf{c},
\label{eq:matrix-twisted-PTdnls}
\end{equation}
where $\mathbf{c}=\left(c_{1},c_{2},\dots,c_{N}\right)^{T}$,
$G=\mathrm{diag}(\gamma_{1}$,$\gamma_{2},\dots,\gamma_{N})$, and 
$N\left(\mathbf{c}\right)=\mathrm{diag}\left(|c_{1}|^{2},|c_{2}|^{2},\dots,|c_{N}|^{2}\right)$
which describe the dissipation and the nonlinear part of the system, and $\mathcal{K}$ is the coupling matrix
$$\mathcal{K}=\begin{pmatrix}0 & e^{-i\phi} & 0 &  &  & e^{i\phi}\\
e^{i\phi} & 0 & e^{-i\phi}\\
0 & e^{i\phi} & 0 & \ddots\\
 &  & \ddots & \ddots & \ddots\\
 &  &  & \ddots & 0 & e^{-i\phi}\\
e^{-i\phi} &  &  &  & e^{i\phi} & 0
\end{pmatrix}_{N\times N}.$$

\begin{figure}[h!]
    \centering
         \includegraphics[width=0.45\textwidth]{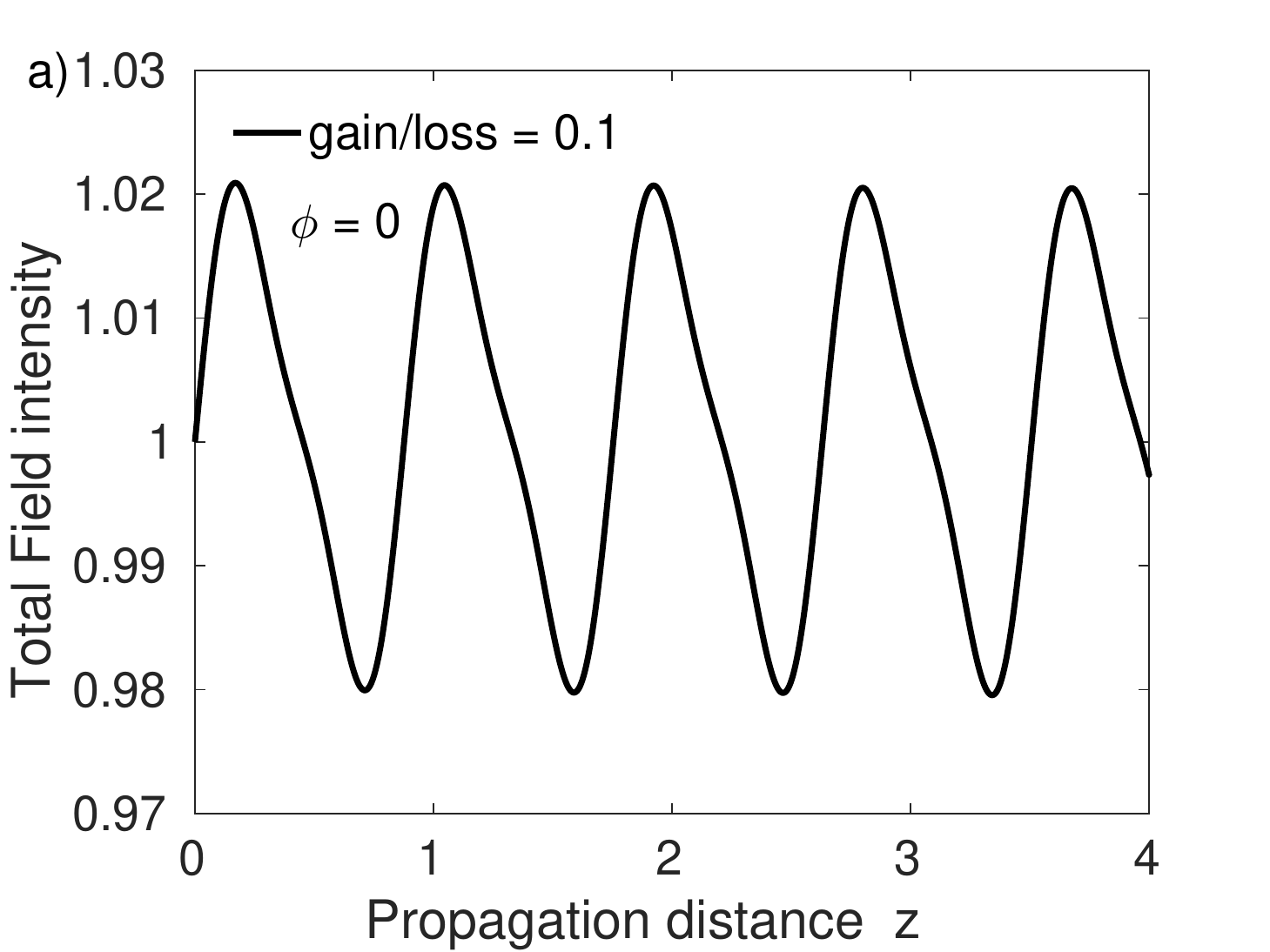}
         \includegraphics[width=0.45\textwidth]{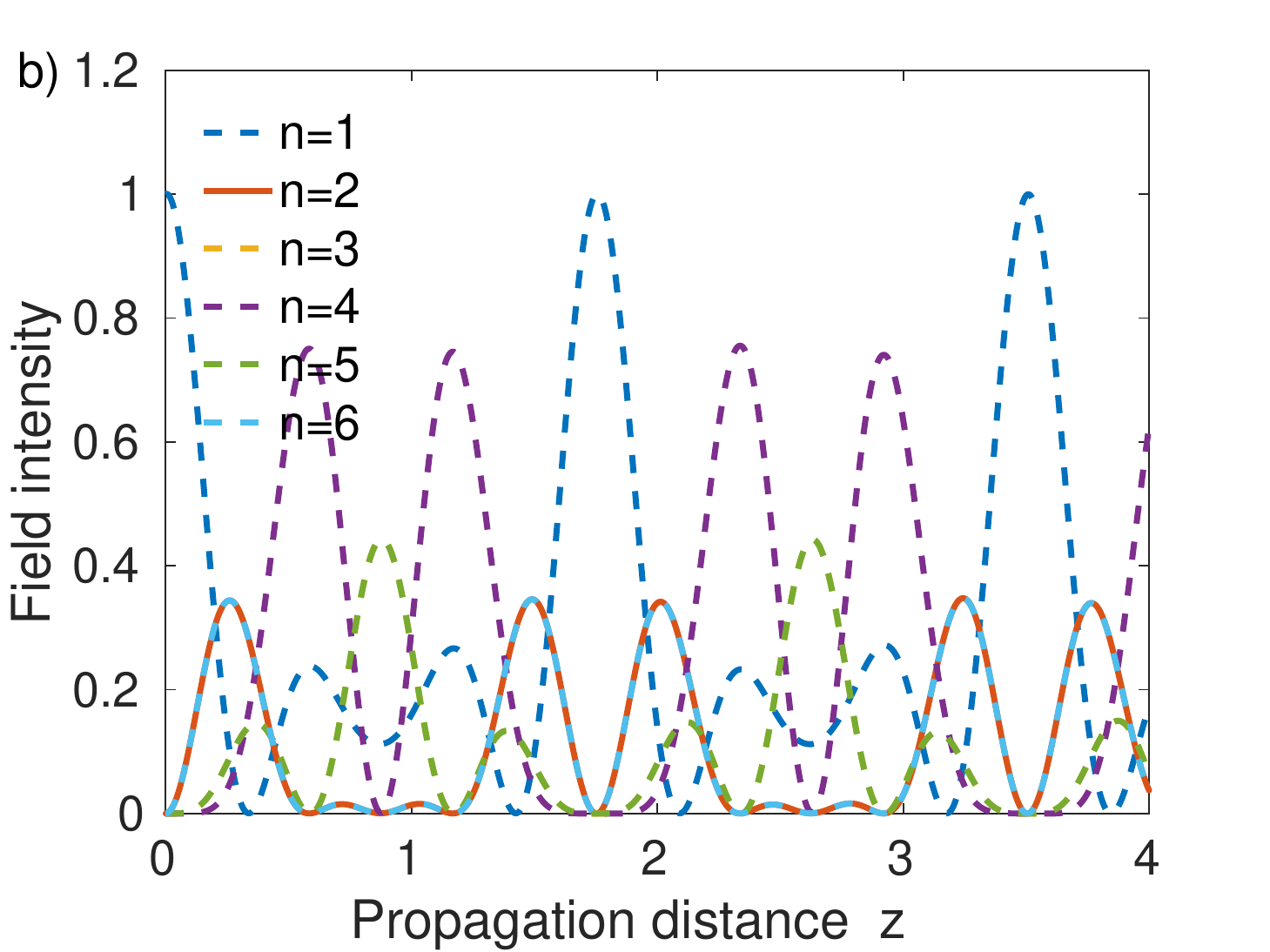}\\
         \includegraphics[width=0.45\textwidth]{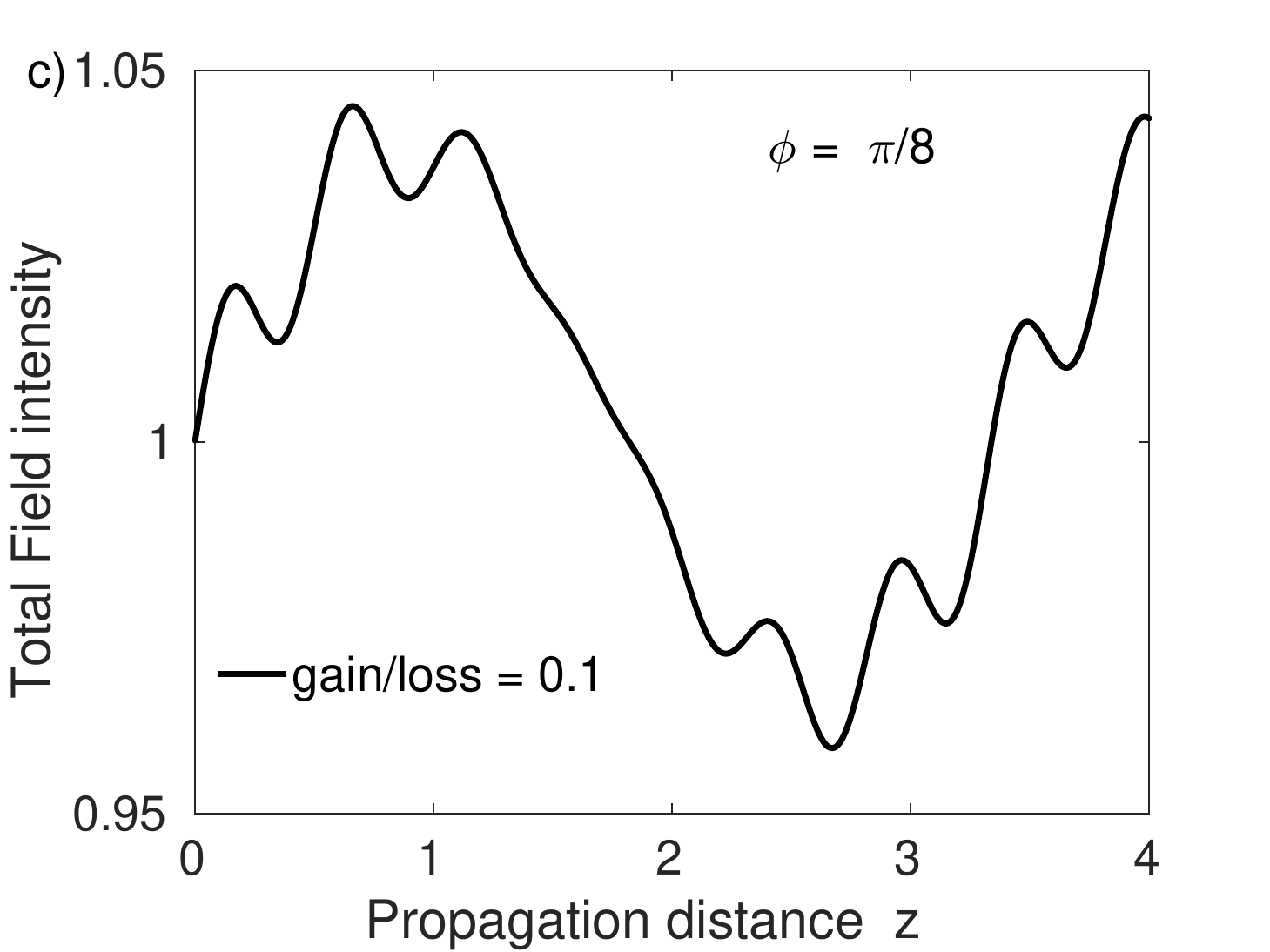}
         \includegraphics[width=0.45\textwidth]{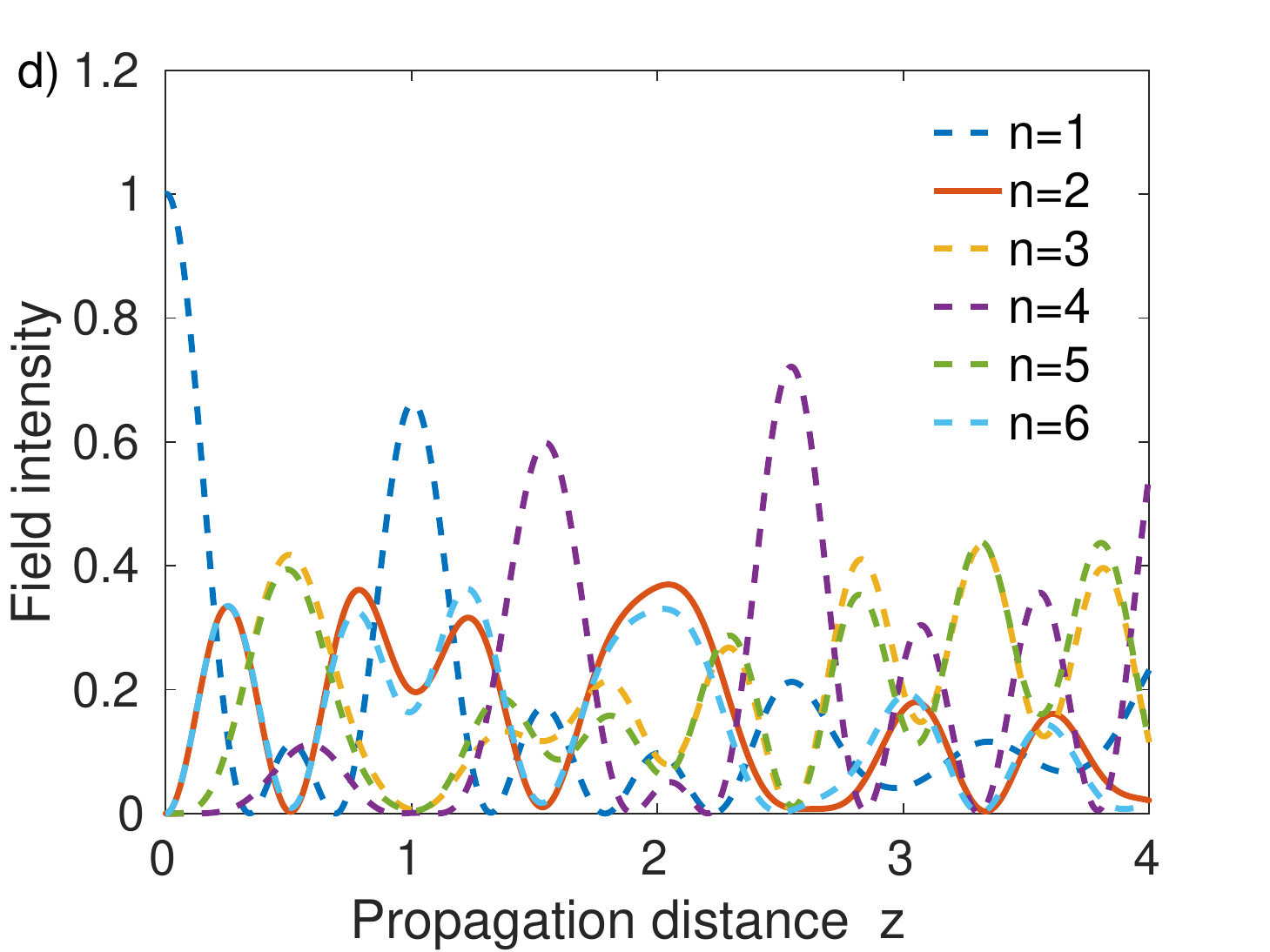}
    \caption{(Color online) Evolution of total power (left), and individual power for cores 1-6 (right) in the nonlinear case 
    with alternating gain-loss in the defocusing regime $\delta d=-1$ with  $\gamma=0.1$. Untwisted and twisted scenarios depicted on 
    top and bottom panels respectively.}
    \label{fig:6cores-Nonlinear-alternating-defocusing}
\end{figure}

At this time we will only present our recent result on the existence of localized modes in the conservative case.
For this, consider the conservative variant of the twisted fiber model, where $\gamma_n=0$. In that case, the equations read:
\begin{equation*}
i\frac{{dc_{n}}}{dz}=k\left( e^{-i\phi}c_{n+1}+e^{i\phi}c_{n-1}\right)+\delta d|c_{n}|^{2}c_{n},
\end{equation*}
\noindent
with periodic boundary conditions $c_{n+N}=c_n$.
If we make the transformation $c_n = a_n(z) e^{i\varphi_n}, \varphi_{n+1}-\varphi_n = \phi$, the equations for $a_n$ are those 
of the classical uniform array, but now the periodic boundary condition is modified to $a_{n+N} = a_n e^{i N\phi}$. 
This transformation also shows in the Hamiltonian
\begin{equation*}
H=\sum_{n=1}^N k \left( c_{n + 1} c_n^*e^{-i \phi} + c_n c^*_{n+1} e^{i \phi}\right)+\frac{\delta d}{2}|c_n|^4 = 
  \sum_{n=1}^N k \left( a_{n + 1} a_n^* + a_n a^*_{n+1} \right)+\frac{\delta d}{2}|a_n|^4. 
\end{equation*}
In the conservative case, the total power $P=\sum_{n=1}^N |c_n|^2$ is also an invariant. Interestingly, in the general case the evolution of $P$ does not depend explicitly on the phase $\phi$,
$$ \frac{d}{dz} \sum_{n=1}^N |c_n|^2 = \sum_{n=1}^N 2\gamma_n |c_n|^2.$$ 
One can see that for small $\gamma_n$ the power is slowy varying and our numerics suggests an oscillatory pattern 
(see figure \ref{fig:6cores-Nonlinear-alternating-defocusing}). While we have not determined a critical value 
of gain/loss for symmetry breaking dynamics, one can speculate that as in the case of the coupler, if at some point all energies propagate in the active 
(gain) fibers, then $\frac{dP}{dz} >0$. Where the phase $\phi$ plays a role not explicitly seen in the dynamics of $P$ is how this transition to dominant 
flow in the fibers with gain is determined by the twist. We are currently exploring such transition.

As it is the case with the untwisted multi-core system with no loss/gain, localized modes extend to the case where there is twist. One can find them by use of asymptotics, so that to leading order, by  assuming that the peak intensity is in fiber $n=1$ and in the case $N=6$
\begin{eqnarray*}
 i\frac{{dc_{1}^{(0)}}}{dz} &=& \delta d |c_1^{(0)}|^2 c_1^{(0)}, \\
 i\frac{{dc_{2}^{(0)}}}{dz} &=& ke^{i\phi} c_1^{(0)}, \\
 i\frac{{dc_{6}^{(0)}}}{dz} &=& ke^{-i\phi} c_1^{(0)}, \\
 i\frac{{dc_{3}^{(0)}}}{dz} &=& ke^{i\phi} c_2^{(0)}, \\
 i\frac{{dc_{5}^{(0)}}}{dz} &=& ke^{-i\phi} c_6^{(0)}, \\
 i\frac{{dc_{4}^{(0)}}}{dz} &=& ke^{i\phi} c_3^{(0)} +ke^{-i\phi} c_5^{(0)},
\end{eqnarray*}
with solutions 
\begin{eqnarray*}
c_{1}^{(0)} &=& \rho e^{-i\delta d \rho^2 z},\\
c_{2}^{(0)} &=& \frac{k}{\delta d \rho} e^{-i\delta d \rho^2 z + i\phi}, \\
c_{6}^{(0)} &=& \frac{k}{\delta d \rho} e^{-i\epsilon_{s}d \rho^2 z - i\phi}, \\
c_{3}^{(0)} &=& \frac{k^2}{(\delta d)^2 \rho^3} e^{-i\delta d \rho^2 z + 2i\phi},\\
c_{5}^{(0)} &=& \frac{k^2}{(\delta d)^2 \rho^3} e^{-i\delta d \rho^2 z - 2i\phi},\\
c_{4}^{(0)} &=& \frac{2k^3}{(\delta d)^3 \rho^5} \cos{3\phi}e^{-i\delta d \rho^2 z},
\end{eqnarray*}
\noindent
where $\rho \gg 1$. 
\noindent
Observe the particular amplitude dependence of $c_4(0)$ on the twist induced phase $\phi$. 
For example, if $\phi = ( 2 m + 1 )\pi/6$, then $c_4(0)=0$.

\begin{figure}[ht!]
    \centering
         \includegraphics[width=0.45\textwidth]{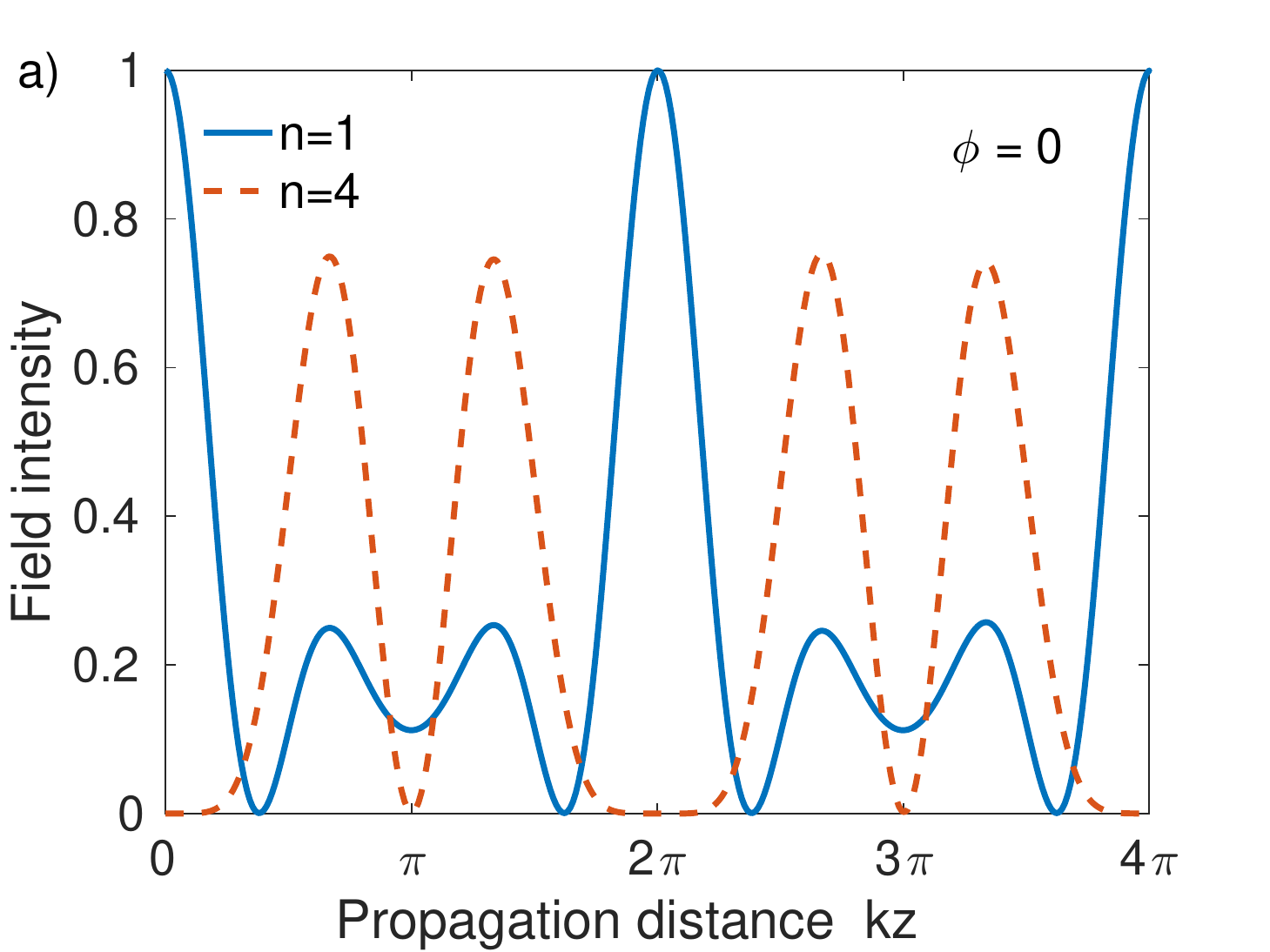}
         \includegraphics[width=0.45\textwidth]{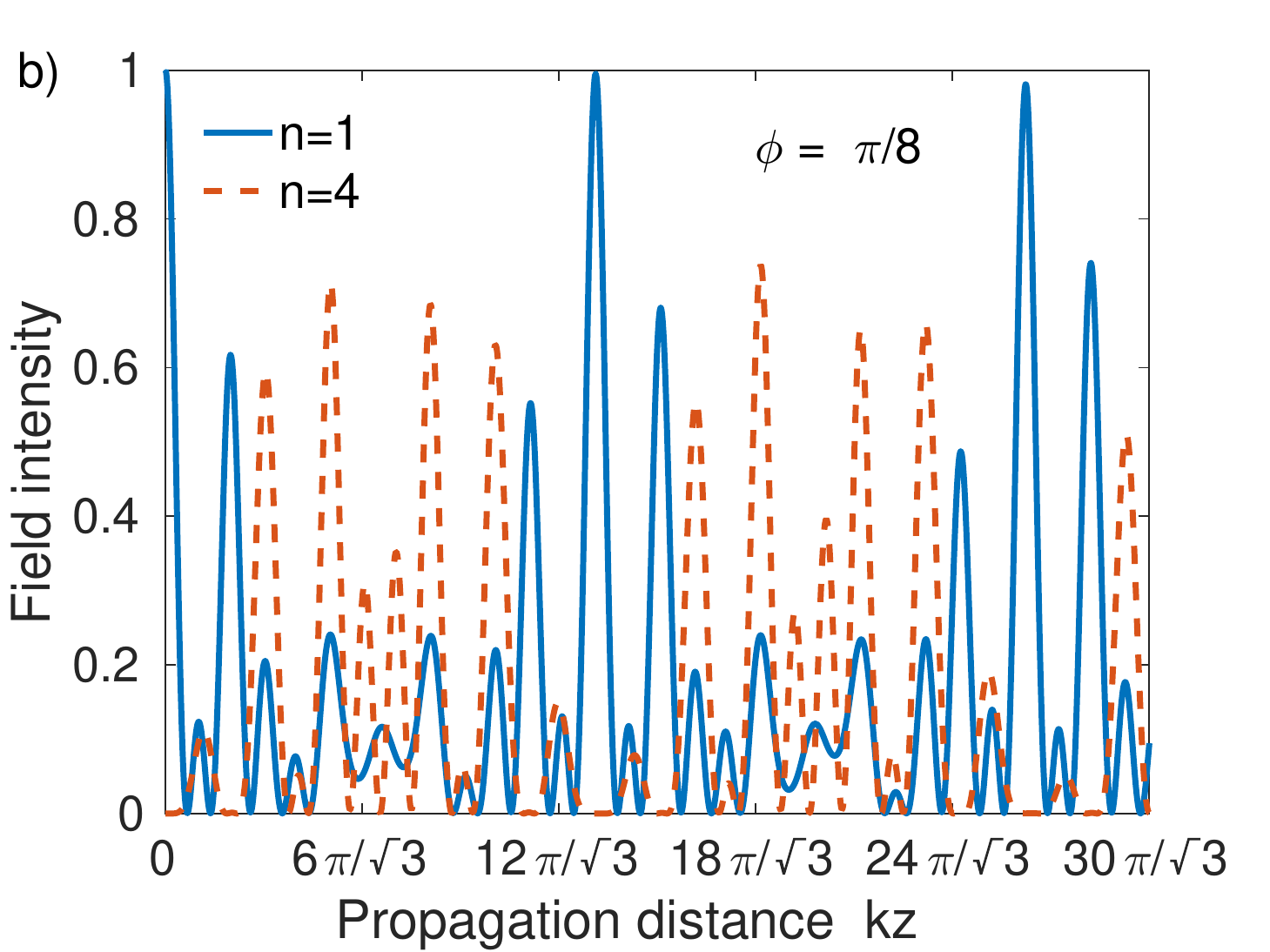}
    \caption{(Color online) Evolution of field intensities for cores 1 (solid) and 4 (dashed) in the nonlinear, conservative, 
    and defocusing case with $\gamma=0$. Left and right panels correspond to non-twisted and twisted scenarios respectively.}
    \label{fig:6cores-Nonlinear-conservative}
\end{figure}

We now present numerical results obtained by integration of Eq. (\ref{eq:twisted-PT-DNLS}).
First we consider a conservative case with $\gamma_{n}=0$ for the defocusing case,
$\delta d=-1$, and parameters $n_{s}=1.552$, 
$R_{0}=8\mu m$, $\lambda=980nm$, $r_{0}=0.91R_{0}$,$k = 3.59cm^{-1}$, and $\gamma=2.05$. The fiber is initially excited in core 1.
Results  in Fig. \ref{fig:6cores-Nonlinear-conservative} present the amplitudes of cores 1 and 4 which clearly highlights the role of the twist in a dramatic 
fashion. On the other hand, this behavior is reminiscent of the linear model treated in Ref. \refcite{8-longhiPTphase2016} suggesting that
the topological feature of the twist and its effect on the dynamics is quite robust to nonlinear contributions which mainly affect the individual phases.

Considering instead the nonlinear CMEs for the six-core fiber with alternating gain and loss $\gamma_{n}=(-1)^{n+1}\gamma$, 
we present numerical results of the integration of Eq. (\ref{eq:twisted-PT-DNLS}) in Fig. \ref{fig:6cores-Nonlinear-alternating-defocusing} for 
a defocusing case. The figure cointains un(twisted) cases in top(bottom) panels. Values used were $n_{s}=1.552$, 
$R_{0}=8\mu m$, $\lambda=980nm$, $r_{0}=0.91R_{0}$, $\gamma=0.1$, $k = 3.59cm^{-1}$, and $\delta d=-1$. As in the conservative case, 
only core 1 is initially excited.
We observe there is an oscillatory behavior of the total power in the absence of twist,
oscillating within a small window of the expected value of the conserved case. Similarly for the twisted case
$\phi=\pi/8.$

In summary, the cases considered are examples of how the induced geometrical twist along with
nonlinearity can cause power weakening/intensification in \pt-symmetric fibers. 
It would be of interest to extend this work for a large array and for concentric multi-core arrays.

\section{Discussion} \label{sec:discussion}

We have discussed some case examples of \pt-symmetric systems in the context of examples stemming from optical waveguide arrays.
For a weakly nonlinear \pt-symmetric dimer, we have found
the governing equations of the slow varying amplitudes of the mode solutions. We 
present two conserved quantities for the nonlinear \pt-symmetric dimer that we used to diagnose the performance of the
numerical integrator. 
For a particular layout of the optical lattice with three waveguides, we show that there exists a parametric transition 
region from unbroken to broken \pt-symmetry.
Of great interest will be to extend these notions to disordered arrays of waveguides.
In the second part of this work we explore how the dynamics of nonlinear fiber with six cores is affected by an induced twist
in the scenario when gain/loss is not present in the model and when there is alternating gain/loss profile. 
The numerical results highlight the potential of inducing a fiber twist to control 
the light dynamics in nonlinear multi-core fibers and suggest a rich scenario for further exploration of parameter space and 
extend this work to for large arrays and arrays with twist ($z$-dependent) management.
Thus, extending present considerations towards identifying localized modes and their dynamical robustness
both in the realm of one dimensional arrays, as well as in higher dimensional settings is one of the natural
directions for future study.

\section*{Acknowledgments}
Castro-Castro acknowledges partial support from LANL grant number LA-UR-15-26399.

\end{document}